\documentclass[10pt,journal,compsoc]{IEEEtran}

\usepackage{balance}

\usepackage[utf8]{inputenc}

\usepackage{graphicx}
\title{1+1$>$2: Programming Know-What and Know-How Knowledge Fusion, Semantic Enrichment and Coherent Application}
\author{Qing~Huang,
        Zhiqiang~Yuan,
        Zhenchang~Xing,
        Zhengkang~Zuo,
        Changjing~Wang,
        Xin~Xia% <-this % stops a space
\IEEEcompsocitemizethanks{\IEEEcompsocthanksitem Q. Huang, Z. Yuan, Z. Zuo, C. Wang are with School of Computer Information Engineering, Jiangxi Normal University, China.\protect
\IEEEcompsocthanksitem Q. Huang and Z. Yuan are co-first authors, Z. Zuo is the corresponding author(zuo803@jxnu.edu.cn)
\IEEEcompsocthanksitem Z. Xing is with Australian National University Australia.
\IEEEcompsocthanksitem X. Xin is with HuaWei.}% <-this % stops an unwanted space
}

\date{August 2021}
\usepackage{graphicx}
\usepackage{float}
\usepackage{array}
\usepackage{tabularx}
\usepackage{multirow}
\usepackage{amsmath}
\usepackage{amssymb}
\usepackage{threeparttable}
\usepackage{setspace}
\usepackage{color}
\usepackage{colortbl}
\usepackage{xcolor}
\usepackage{soul}
\usepackage{booktabs}
\usepackage{makecell}
\usepackage{algorithm} 
\usepackage{algpseudocode}
\usepackage{hyperref}
\usepackage{caption}

\usepackage{verbatim}
\usepackage{picins}

\definecolor{mygray}{gray}{.9}
\definecolor{mypink}{rgb}{.99,.91,.95}
\definecolor{mycyan}{cmyk}{.3,0,0,0}
\definecolor{myyellow}{RGB}{255,230,204}
\definecolor{mybule}{RGB}{218,232,252}
\definecolor{mygreen}{RGB}{213,232,212}
\definecolor{titleColor}{RGB}{102,102,102}

\begin{document}
\IEEEtitleabstractindextext{%
\begin{abstract}
Software programming requires both API reference (know-what) knowledge and programming task (know-how) knowledge. 
Lots of programming know-what and know-how knowledge is documented in text, for example, API reference documentation and programming tutorials.
To improve knowledge accessibility and usage, several recent studies use Natural Language Processing (NLP) methods to construct API know-what knowledge graph (API-KG) and programming task know-how knowledge graph (Task-KG) from software documentation. 
Although being promising, current API-KG and Task-KG are independent of each other, and thus are void of inherent connections between the two types of knowledge.
Our empirical study on Stack Overflow questions confirms that only 36\% of the API usage problems can be answered by the know-how or the know-what knowledge alone, while the rest questions requires a fusion of both.
Inspired by this observation, we make the first attempt to fuse API-KG and Task-KG by API entity linking.
This fusion creates nine categories of API semantic relations and two types of task semantic relations which are not present in the stand-alone API-KG or Task-KG.
According to the definitions of these new API and task semantic relations, our approach dives deeper than surface-level API linking of API-KG and Task-KG, and infer nine categories of API semantic relations from task descriptions and two types of task semantic relations with the assistance of API-KG, which enrich the declaration or syntactic relations in the current API-KG and Task-KG.
Our fused and semantically-enriched API-Task KG supports coherent API/Task-centric knowledge search by text or code queries.
We have implemented our approach on Java programming documentation and built a web tool to search and explore API and programming task knowledge.
Our evaluation confirms the high-accuracy of our knowledge extraction, fusion and enrichment methods, and the effectiveness and usefulness of our API-Task KG for answering Stack Overflow questions.
\end{abstract}

% Note that keywords are not normally used for peerreview papers.
}

% make the title area
\maketitle
\IEEEdisplaynontitleabstractindextext
\IEEEpeerreviewmaketitle

\IEEEraisesectionheading{\section{INTRODUCTION}}
To use certain APIs in programming tasks, developers resort to software documentation, such as API reference and programming tutorials.
API reference documentation explains the functionalities and usage directives of individual APIs, while programming tutorials describe different tasks and scenarios in which some APIs are (or are not) applicable and how to use them properly or avoid misuses.
In the literature, the former is referred to as know-what, while the latter is know-how.
Recent studies construct API knowledge graph (API-KG) \cite{Li2018ImprovingAC,Liu2019GeneratingQC, Ren2020APIMisuseDD} and programming task knowledge graph (Task-KG) \cite{Sun2019KnowHowIP,Sun2020TaskOrientedAU} from software documentation.
They show that lifting programming knowledge from semi- or unstructured text into structured knowledge enables API- or task-centric knowledge recommendation and can proactively reduce API misuses.

\begin{figure}
    \centering 
    \includegraphics[width=0.47\textwidth]{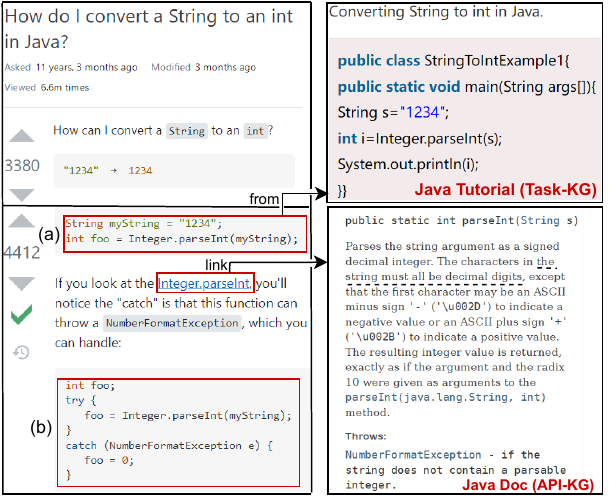}
    \caption{Need for the Fusion of Programming Know-What and Know-How Knowledge 
\vspace{-6mm}
} \label{fig:movitation}
   
\end{figure}

Although these studies show the promising of knowledge graph methods for software engineering, the two types of knowledge are currently independently analyzed, extracted and used. 
We conduct an empirical study of 100 randomly-selected API usage questions on Stack Overflow (SO), and find that only 36\% of the API usage questions can be answered by know-what or know-how knowledge alone, while the rest requires a fusion of both.
For example, Figure~\ref{fig:movitation} shows a highly up-voted and frequently-viewed question ``How do I convert a String into an int in Java?\footnote{\href{https://stackoverflow.com/questions/5585779/how-do-i-convert-a-string-to-an-int-in-java}{https://stackoverflow.com/questions/5585779}}''.
To answer this question, the accepted answer posts a code example (as shown in Figure~\ref{fig:movitation}-a) which comes from a converting-string-to-int task in the Java tutorial~\footnote{\href{https://www.javatpoint.com/java-string-to-int}{https://www.javatpoint.com/java-string-to-int}}(i.e., know-how knowledge in Task-KG).
At this time, while learning about \textit{Integer.parseInt()} from Java API specification (i.e., know-what knowledge in API-KG), we find a flaw in the code example which will throw a \textit{NumberFormatException} if the string does not contain a parsable integer. 
This flaw is also mentioned in the questions~\footnote{\href{https://stackoverflow.com/questions/18711896/how-can-i-prevent-java-lang-numberformatexception-for-input-string-n-a}{https://stackoverflow.com/questions/18711896}}\footnote{ \href{https://stackoverflow.com/questions/39849984/what-is-a-numberformatexception-and-how-can-i-fix-it}{https://stackoverflow.com/questions/39849984}} on SO.
Inspired by this, the code example is optimized with a “try-catch” structure (as shown in  Figure~\ref{fig:movitation}-b).
This illustrates the necessity of fusing know-what and know-how knowledge.
Know-what knowledge in API-KG enables developers to learn about the API and optimize the code example used in know-how knowledge in Task-KG; in turn, know-how knowledge complements know-what knowledge by guiding developers to use the API in programming tasks with code examples.

In addition, we find that when know-what and know-how knowledge are fused into a whole, the whole will be more than the sum of the two parts.
Specifically, from the empirical study, we explore nine categories of API semantic relations (e.g., behavior difference, function collaboration) and two types of task semantic relations (task-align and task-overlap). 
The relevant examples are shown in section~\ref{sec:empiricalstudy}. 

Inspired by our empirical study, we propose a novel approach to construct a fused and semantically-enriched API-Task KG on the basis of independent API-KG and Task-KG.
Our approach consists of four steps: KG construction, KG fusion, KG enrichment and KG application.
For KG construction, we integrate the methods in~\cite{Li2018ImprovingAC,Liu2019GeneratingQC}
to construct the API-KG from API reference documentation and use the method in~\cite{Sun2019KnowHowIP} to construct the Task-KG from programming tutorials.
API-KG represents API entities, API function and directive sentences and API declaration relations, and Task-KG represents task entities (as $<$action, object$>$ tuples), task attributes (e.g., code examples, code summary) and task dependencies.
For KG fusion, we design an API-entity linking method to fuse API-KG and Task-KG, which is the combination at the entity level rather than the text level.
For KG enrichment, according to the definition of the API/task semantic relations in our empirical study, we enrich API-KG with nine types of API semantic relations inferred from task descriptions in Task-KG, and enrich Task-KG with two types of task semantic relations inferred from the APIs used in the tasks and API relations in API-KG.
For KG application, we develop an API- and task-centric search engine and a web tool for searching and exploring programming know-what and know-how knowledge by text or code queries.

Our evaluation focuses on two aspects: the intrinsic quality of the key steps of our approach, and the effectiveness and usefulness of our API-Task KG.
The key steps of our approach include function sentence identification, directive sentence identification, task phrase extraction and API-packet extraction, inference of API semantic relations, and inference of task semantic relations.
Considering the sheer amount of data instances in KG, we use a statistical sampling method~\cite{Singh1996ElementsOS} to estimate the accuracy of these steps. 
Our analysis shows: 94.30\% and 92.70\% accuracy for function and directive sentence identification; 89.84\% F1-score for task phrase extraction and 89.72\% accuracy for API-packet extraction; 86.61\% accuracy for KG fusion; 97.27\% and 90.55\% average accuracy for API semantic relation inference and task semantic relation inference, respectively.

To evaluate the effectiveness and usefulness of our API-Task KG, we sample 30 frequently-viewed Stack Overflow (SO) questions from 100 questions in our empirical study, and collect the accepted answers of these questions. 
We use either the question title or the code snippet in the question to search our API-Task KG in our prototype tool, and compare the recommended API and/or task knowledge with the accepted answers. 
For 23 out of 30 questions, the recommended knowledge includes the accepted answers. 
In addition, the API-Task linkages and the inferred API semantic relations and task semantic relations allow our tool to suggest rich extended knowledge for the question, which are not covered in the accepted answer. 
To verify whether the extended suggestion is helpful or results in unnecessary information overload, we conduct a user study in which two groups of graduate students (6 each) use either our API-Task KG tool or independent API-KG and Task-KG tools to find the solutions to the 30 SO questions. 
Our study suggests that our API-Task KG helps the students reduce completion time by 27.5\% and improve accuracy by 15.0\%, and the extended knowledge inspires the student to explore new ideas or even come up with better solutions.

Our idea can be regarded as an instance of the general phenomenon ``people always distill or infer richer knowledge by fusing the independent knowledge, and leverage new knowledge to solve the issues not covered by the independent knowledge.''~\cite{bagherzadeh2021competing}.
The contributions are as follows:

\begin{itemize}
    \item Through an empirical study, we understand the need of the fusion of programming know-what and know-how knowledge, and infer fine-grained semantic relations including nine categories of API semantic relations and two types of task semantic relations.

    \item To our best knowledge, we are the first to investigate the fusion and cross-enrichment of programming know-what and know-how knowledge, and the application of fused knowledge in API search.

    \item We propose a systematic approach for KG construction, fusion and semantic enrichment, and build a prototype tool to search programming know-what and know-how knowledge coherently.

    \item We apply our approach to Java API reference and programming tutorials. Our evaluation confirms the high quality of the resulting API-Task KG and provide the evidence of its effectiveness and usefulness. 
    Our data package can be found here\footnote{\href{https://anonymous.4open.science/r/know-what-and-know-how-data-223B}{know-what-and-know-how-data}}

\end{itemize}

\vspace{-3.5mm}
\section{EMPIRICAL STUDY} 
\label{sec:empiricalstudy}

%\zq{The entire section}
API-KG and Task-KG contain an abundance of structural knowledge for solving programming issues. 
To understand the role of know-what knowledge in API-KG and know-how knowledge in Task-KG in addressing API usage questions,
we conduct an empirical study on the questions selected from Stack Overflow, and answer the following research questions: \textbf{RQ1} - \textit{If an API usage question can be solved using API-KG or Task-KG alone, we regard it as Q1; otherwise we regard it as Q2. What are the percentages of Q1 and Q2, respectively?} \textbf{RQ2} - \textit{Does Q1 imply the semantic relations that are not in API-KG and Task-KG?}
\textbf{RQ3} - \textit{Does Q2 imply the semantic relations that are not in API-KG and Task-KG?}

\vspace{-3mm}
\subsection{Study Design}

\subsubsection{Background of API-KG and Task-KG}\label{sec:Background}
API-KG~\cite{Li2018ImprovingAC, Liu2019GeneratingQC}, and Task-KG~\cite{Sun2019KnowHowIP,Sun2020TaskOrientedAU}, have been proposed to improve the know-what and know-how knowledge accessibility by lifting programming knowledge from semi- or unstructured text into structured knowledge. 
This structured knowledge exists in the form of relations of KG.
Since these relations are extracted from the document structure, we refer to them as declaration (i.e., syntactic) relations.
The API-KG contains eight declaration relations, i.e., \textit{contain}, \textit{extend}, \textit{implement}, \textit{throw}, \textit{hasMethod}, \textit{hasParameter}, \textit{hasField} and \textit{hasConstructor}.
The Task-KG contains three declaration relations, i.e., \textit{parent-child}, \textit{sibling} and \textit{temporal}.
However, the know-what knowledge and know-how knowledge reflected in these declarations are independent, and we argue that this is not enough to solve programming issues.
% However, the know-what knowledge and know-how knowledge reflected in these declaration relations are independent of each other and are not enough to solve programming issues.
Therefore, we conduct an empirical study to understand the complementary nature of know-what and know-how knowledge and explore the opportunity to combine them for facilitating the discovery of more fine-grained semantic relations.

\vspace{-2mm}
\subsubsection{Data Preparation}

We select API usage questions tagged with ``java'' in the SO data dump~\cite{StackOverflow}, and  filter them based on two criteria: 1) the question title begins with ``how to'', or contains the API name; and 2) there is an accepted answer with sufficient details rather than just referencing materials.
We obtain 8,667 questions, and sort them in descending order by view count.
To minimize manual effort, we select only 100 questions for this empirical study.
Note that we make sure that the 100 questions cover as many topics as possible.
Specifically, there are ten topics: \textit{Base Type}, \textit{Collections}, \textit{Data}, \textit{Date}, \textit{File}, \textit{IO}, \textit{Network}, \textit{Reflection}, \textit{Swing}, \textit{Thread}.
For example, the topic \textit{Base Type} indicates that the selected questions are primarily about String (or Int) operations.
Given these topics, we arrange for two master's students with more than four years of Java development experience to independently select 100 questions from the sorted questions, with a Cohen’s Kappa agreement~\cite{Landis1977AnAO} of 0.842, i.e., almost perfect agreement.
Finally, we obtain 10 questions for \textit{Base Type}, 30 for \textit{Collections}, 7 for \textit{Data}, 9 for \textit{Date}, 7 for \textit{File}, 8 for \textit{IO}, 8 for \textit{Network}, 6 for \textit{Reflection}, 7 for \textit{Swing}, and 8 for \textit{Thread}.

As developers tend to answer the questions on Stack Overflow with API descriptions (e.g., API function and directive sentences~\cite{Liu2019GeneratingQC}) and code examples, we consider them as answer points following the treatment mentioned in~\cite{Liu2020GeneratingCB}. 
Note that an accepted answer may contain more than one answer point, but it has only one main answer point (the most direct answer to the question), and all of the rest are secondary answer points.

Given the 100 selected questions, we ask those two Master 
students to extract answer points from the accepted answer of each question independently. 
Since two students might extract different answer points to the same question, we assign a PhD student (with more than six years of Java development experience) to check the different answer points and solve their conflicts. 
The Cohen’s Kappa coefficient between two Master students is 0.880 (i.e., almost perfect agreement).
Finally, we obtain 206 answer points, including 100 main answer points.
As a result, we collect 100 top-viewed questions, each of which has a main answer point and optionally some secondary answer points.

\vspace{-2.5mm}
\subsubsection{Protocol}
To answer RQ1, we give these two students a 10-minute training session on how to distinguish Q1 from Q2 in the 100 top-viewed questions.
If the main answer point to each question could be retrieved in API-KG or Task-KG separately, the question belongs to Q1; otherwise it belongs to Q2.
However, considering that there are differences between the descriptions in the answer point and the ones in API-KG or Task-KG, those two Master students would consider the semantic similarity between the two descriptions, rather than literal similarity, when retrieving answer points.
For example, the answer point to the question ``\href{https://stackoverflow.com/questions/355089/difference-between-stringbuilder-and-stringbuffer/20512746#20512746}{Difference between StringBuilder and StringBuffer''} is ``StringBuffer is synchronized, StringBuilder is not.'', while the description in API-KG is ``If synchronization is required then it is recommended that StringBuffer be used''.
These two descriptions are literally-different but semantically-similar.
For the answer point annotated differently, the PhD student serves as an arbiter to make the final decision.

To answer RQ2 and RQ3, we give those two Master students a 20-minute training session on telling them the existing declaration relations (i.e.,  syntactic relations) in API-KG and Task-KG, so that they determine whether or not semantic relations that are not in API-KG and Task-KG exist in the questions of Q1 or Q2 type.

\vspace{-4.0mm}
\subsection{Result and Analysis}
Through analysis of the 100 questions and corresponding 206 answer points, we obtained the answers to the three research questions.
For RQ1, the percentages of Q1 and Q2 are 36\% and 64\%, respectively.
For RQ2, we find four implicit semantic relations, i.e., Function Similarity (23.5\%), Function Opposite (9.8\%), Behavior Difference (21.6\%) and Function Replace (45.1\%).
For RQ3, we find seven implicit semantic relations, i.e., Function Collaboration (14.3\%), Type Conversion (12.5\%), Implement Constraint (10.7\%), Logic Constraint (7.1\%), Efficiency Comparison (5.3\%) , Task Align (33.9\%) and Task Overlap (16.1\%).

\vspace{-3.0mm}
\subsubsection{Answer to RQ1}
Given each question, two students search for its main answer point in API-KG~\cite{Liu2019GeneratingQC} and Task-KG~\cite{Sun2019KnowHowIP} separately, and annotate the question as Q1 or Q2 depending on whether the answer is retrieved.
Statistically, Q1 accounts for 36\% while Q2 for 64\%. 
Finally, we calculate the agreement between the two students for annotating, and the Cohen’s Kappa coefficient is 0.864 (i.e., almost perfect agreement).

\vspace{-3mm}
\subsubsection{Answer to RQ2} \label{Answer for RQ2}

We analyze if the semantic relations that are not in  API-KG and Task-KG exist in Q1, and find that 78.3\% of the answer points in Q1 imply 4 types of new API semantic relations.

\textbf{Function Similarity} relation is defined that two API entities have similar usage. 
We find that 23.5\% of the answer points in Q1 questions reveal this API relation.
For the question ``\href{https://stackoverflow.com/questions/1938855/how-to-store-a-large-10-digits-integer}{How to store a large (10 digits) integer?}'', the answer point ``If at any point you need bigger numbers, you can try java.math.BigInteger, or java.math.BigDecimal'' reveals that both \textit{java.math.BigInteger} and \textit{java.math.BigDecimal} can be used to operate bigger numbers. 

\textbf{Function Opposite} relation is defined that two API entities have opposite usage. 
We find that 9.8\% of the answer points in Q1 questions reveal this API relation.
For the question ``\href{https://stackoverflow.com/questions/19960243}{How to store IP Address range vs location}'', the answer point ``The higherEntry(K key) does the opposite of the lowerEntry(K key), meaning higherEntry(K key) returns a key-value mapping associated with the least key strictly greater than the given key, or null if there is no such key.'' reveals that \textit{higherEntry(K key)} and \textit{lowerEntry(K key)} have the opposite function when operating the element.

\textbf{Behavior Difference} relation is defined that two similar API entities behave differently when completing the same task. 
We find that 21.6\% of the answer points in Q1 questions reveal this API relation.
For the question ``\href{https://stackoverflow.com/questions/2703984/what-is-the-difference-between-the-add-and-offer-methods-in-a-queue-in-java}{What is the difference between the add and offer methods in a Queue in Java}'', the answer point ``when element can not be added to collection the add method throws an exception and offer doesn't.'' reveals that \textit{add()} and \textit{offer()} behave differently when adding elements to a collection fails.

\textbf{Function Replace} relation is defined that one API entity should be replaced by another API entity in some specific condition.
We find that 45.1\% of the answer points in Q1 questions reveal this API relation.
For the question ``\href{https://stackoverflow.com/questions/2772836}{Iterator has.next() - is there a way to get the previous element instead of the next one?}'', the answer point ``You can use ListIterator instead of Iterator. '' reveals that developers should use \textit{ListIterator} and \textit{Iterator} when you need to get the previous element.
This is because \textit{ListIterator} has \textit{previous()} and \textit{hasPrevious()} methods.

In this process, we calculate the agreement between the two students for their annotations, and the Cohen’s Kappa coefficient is 0.836 (i.e., substantial agreement).

\vspace{-3.0mm}
\subsubsection{Answer to RQ3}\label{empir:RQ3}

We analyze if the semantic relations that are
not in API-KG and Task-KG exist in Q2, and
find that 79.3\% of the answer points in Q2 imply 5
types of new API semantic relations and 2 kinds of task
semantic relations.

\textbf{Function Collaboration} relation is defined that two API entities should be used together when accomplishing a task.
We find that 14.3\% of the answer points in Q2 questions reveal this API relation.
For the question ``\href{https://stackoverflow.com/questions/31955193}{how to start a function after stop typing in a JTextField in java}'', the answer point ``Use a Swing Timer and a DocumentListener, each time the Document is updated, reset the Time'' reveals that developers should use \textit{Time} and \textit{DocumentListener} together when starting a function after stop typing in a JTextField.

\textbf{Type Conversion}
relation is defined that that two API entities can be converted to each other.
We find that 12.5\% of the answer points in Q2 questions reveal this API relation.
For the question ``\href{https://stackoverflow.com/questions/5683728}{Convert java.util.Date to String}'', the answer point ``Convert a Date to a String using format() method'' reveals that \textit{Date} could be converted to \textit{String}. 

\textbf{Implementation Constraint} relation is defined that the implementation of one API relies on the other API.
We find that 10.7\% of the answer points in Q2 questions reveal this API relation.
For the question ``\href{https://stackoverflow.com/questions/9791931/remove-equal-item-from-java-list}{Remove equal item from java list}'', the answer point ``If you properly override the equals method, you can then just use the remove method'' reveals that the implementation of \textit{remove()} relies on \textit{equal()}. 

\textbf{Logic Constraint} relation is defined that one API should be called before or after using another API.
We find that 7.1\% of the answer points in Q2 questions reveal this API relation.
For the question ``\href{https://stackoverflow.com/questions/20666454/how-to-pause-all-running-threads-and-then-resume/20666930#20666930}{How to pause all running threads? and then resume?}'', the answer point ''wait() suspends the current thread until another thread calls notify() to wake up.'' reveals that developers need to use \textit{notify()} to wake up a thread after suspending it with \textit{wait()}.

\textbf{Efficiency Comparison} relation is defined that efficiency comparison of two APIs in some specific conditions.
We find that 5.3\% of the answer points in Q2 questions reveal this API relation.
For the question ``\href{https://stackoverflow.com/questions/10442775/performance-of-treemap-hashmap-and-linkedhashmap}{Performance of TreeMap, HashMap and LinkedHashMap?}'', the answer point ``Use HashMap unless you need for ordering. HashMap is faster.'' reveals that \textit{HashMap} is more efficient than \textit{TreeMap} and \textit{LinkedHashMap} when sorting is not required. 

\textbf{Task Align} relation is defined that different tasks achieve the same goal.
Actually, the same questions are often solved in different ways. 
We find that 33.9\% of the answer points in Q2 questions reveal this API relation.
For the question ``\href{https://stackoverflow.com/questions/326390/how-do-i-create-a-java-string-from-the-contents-of-a-file}{How to create a Java string from the contents of a file?}'', as shown in the accepted answers, the developers offer six ways to answer this question, and each contains a different code example (i.e., different tasks). 
For example, the tasks can be completed with \textit{File.readString()} and \textit{Files.readAllLines()}.

\textbf{Task Overlap} relation is defined that a snippet of code for one task may overlap with a portion of code for another task in terms of API usage.
We find that 16.1\% of the answer points in Q2 questions reveal this API relation.
For the question ``\href{https://stackoverflow.com/questions/157944/create-arraylist-from-array?rq=1}{Create ArrayList from array}'' and the question ``\href{https://stackoverflow.com/questions/1005073/initialization-of-an-arraylist-in-one-line?rq=1}{Initialization of an ArrayList in one line}'', these two different questions are linked to each other by URLs on Stack Overflow. 
This is because some of the code in their accepted answers overlaps, i.e., ``ArrayList$<$String$>$ places = new ArrayList$<$$>$(Arrays.asList());``. 

In this process, we calculate the agreement between the two students for annotation, and the Cohen’s Kappa coefficient is 0.857 (i.e., almost perfect agreement).

%\vspace{1mm}
\noindent\fbox{\begin{minipage}{8.4cm} \emph{Fusion of know-what and know-how knowledge is frequently needed to solve programming issues. This fusion requires nine categories of API semantic relations and two types of task semantic relations, which are not present in the existing independent API-KG and Task-KG.} \end{minipage}}\\

\vspace{-2.0mm}
\section{APPROACH}
\begin{figure}%[H]
    \centering
    \includegraphics[width=0.47\textwidth]{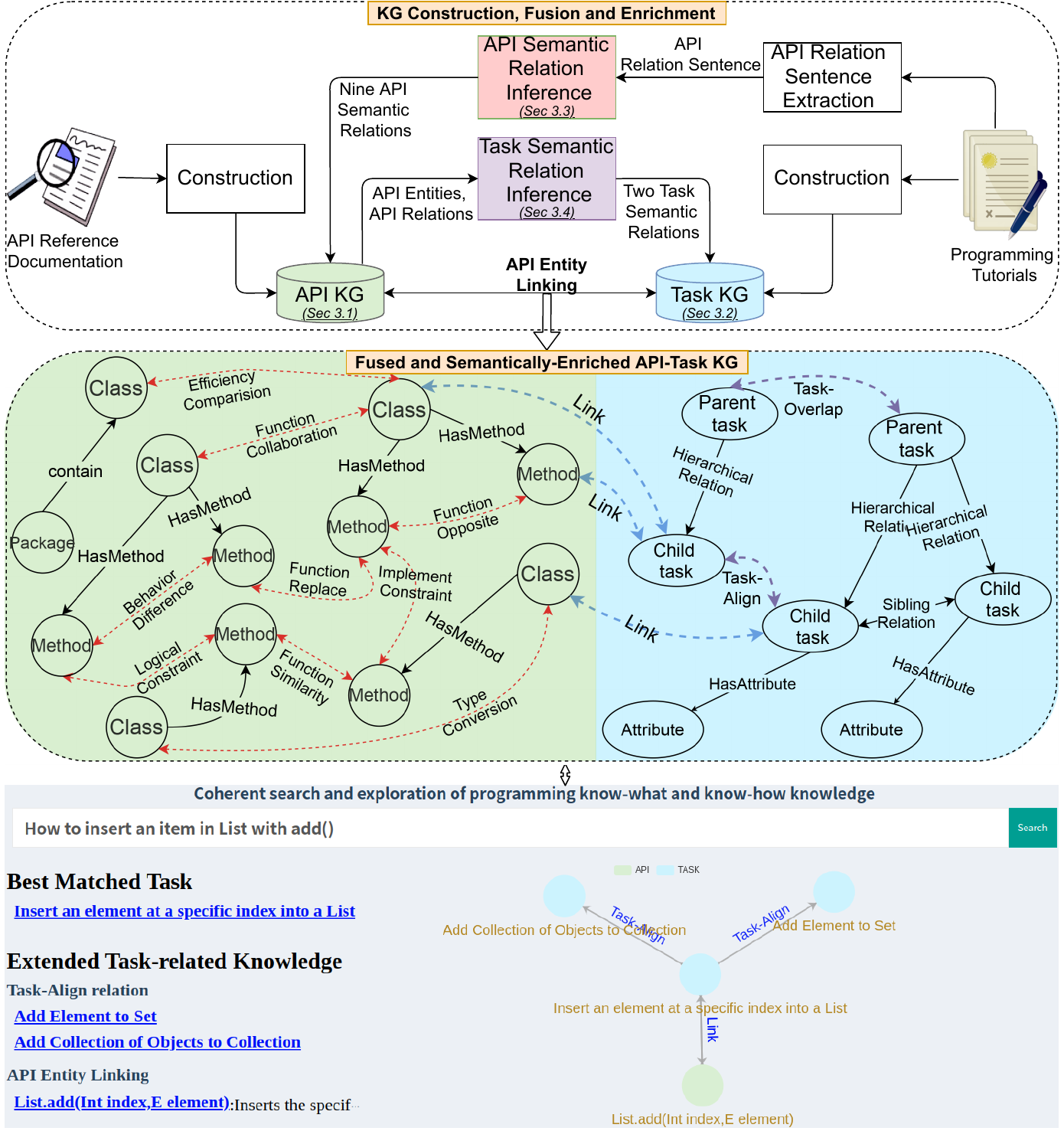}
     \footnotesize
     Note that the figure in the middle, the red and purple dotted line represents the enriched semantic relation, and the blue dotted line represents the entity link between API-KG and Task-KG.
     \vspace{-2.5mm}
    \caption{KG Construction, Enrichment and Application} 
    \label{fig:overview}
    \vspace{-4.7mm}
\end{figure}

Our empirical study suggests the necessity of the fusion and mutual enrichment of know-what and know-how knowledge.
As illustrated in Figure~\ref{fig:overview}, our approach first constructs an API-KG and a Task-KG separately (Section~\ref{sec:kgconstruction}) and links the two KGs by API entities mentioned in task entities (Section~\ref{kg-link}).
Next, our approach infers new semantic relations from one KG to enrich the other KG (Section~\ref{sec:enrich API-kg} and Section~\ref{sec:enrich task-kg}).
Finally, we develop an API/Task-centric search engine and a web interface that recommends programming know-what and know-how knowledge as a unified whole for text or code queries (Section~\ref{sec:kgapplication}). Moreover, the figure in the middle shows an overview of API-Task KG, and the bottom figure shows an application screenshot.

\vspace{-2.5mm}
\subsection{KG Construction}\label{sec:kgconstruction}

We construct API-KG and Task-KG from API reference documentation and programming tutorials respectively.

\vspace{-3.0mm}
\subsubsection{API-KG Construction}\label{sec:api-kg}

We adopt the methods proposed in~\cite{Li2018ImprovingAC, Liu2019GeneratingQC} to construct API-KG. 
These methods assume that API reference documentation follows a consistent semi-structured format, from which we extract eight types of API entities (Package, Class, Interface, Exception, Method, Parameter, Field) and eight types of API declaration relations (contain, extend, implement, throw, hasMethod, hasParameter, hasField, hasConstructor).
These API entities and declaration relations constitute a so-called API skeleton graph~\cite{Li2018ImprovingAC, Liu2019GeneratingQC}.
Then we extract Function or Directive Sentence~\cite{Liu2019GeneratingQC} from API description as API attributes. 
The former defines the functionality of the API, the latter tells how to use the APIs correctly.
% Function Sentence is used to describe the functionality of API entities, while the Directive Sentence is used to describe the correct or incorrect use of API entities.
% Next, we extract two types of sentences (function and directive~\cite{Liu2019GeneratingQC}) from API descriptions as API attributes.
Specifically, we resolve the pronouns in API descriptions with NeuralCoref~\cite{NeuralCoref}, and split the descriptions into sentences with Spacy~\cite{Spacy}.
As API descriptions contain many API tokens, general English tokenizer (e.g, Jieba~\cite{Jieba}) would break an API token into several tokens, which negatively affects sentence splitting. 
For example, the method ``add(index, E)'' will be broken into ``add(index'' and ``E)'' with the general tokenizer.
Therefore, we use software-specific tokenizer as in the previous work~\cite{Li2018ImprovingAC, Liu2019GeneratingQC}.
Then we do part-of-speech (POS) tagging for the sentences and identify the function sentences based on two criteria: (1) API name is the subject in the sentence; or (2) the sentence starts with a verb phrase, for example, ``returns the numbers of the elements in this collection'' for \textit{Collection.size()}.
Meanwhile we identify the directive sentences according to 86 keywords (e.g., must, only) defined in an empirical study of API directives in API documentation \cite{Martin2011WhatSD}.
For example, according to the keyword ``must'', we extract the directive sentence ``All methods on the Array interface \underline{must} be fully implemented if the JDBC driver supports the data type''.
We use the linking methods in~\cite{Li2018ImprovingAC} to link the identified function and directive sentences to the corresponding APIs.
% \zc{If there are multiple function sentences, how do we deal with it? Merge as one function attribute? I think this merge is necessary for later API entity linking?}.
Following the practice in~\cite{Li2018ImprovingAC, Liu2019GeneratingQC}, if there are multiple function sentences, we merge them as one function-sentence attribute.
We keep each directive sentence as a separate API attribute.

\vspace{-3.0mm}
\subsubsection{Task-KG Construction}\label{sec:task-kg}

We follow the method in~\cite{Sun2019KnowHowIP} to construct Task-KG.
In particular, we extract task entities, attributes and relations from the semi-structured programming tutorials.

\textbf{a) Task Entity Extraction.}
% \paragraph{a) Task Entity Extraction}
We use the same set of NLP tools as described in Section~\ref{sec:api-kg} to process task texts, split them into sentences and obtain the POS tags for the sentences.
Different from~\cite{Sun2019KnowHowIP} represents task entities at sentence level, we extract verb phrases as task entities which are more fine-grained and can be more accurately predicted (see Section~\ref{sec:quality task-kg}).
We extract verb phrases from the sentences with Spacy and annotate them independently by two authors if a verb phrase implies a clear task intention and should be considered as a task entity.
For example,``Convert Set to List'' implies a task, while ``pass the set as parameter to addAll()'' does not.
If the two annotators disagree with each other, the third author serves as an arbiter and makes the final decision. 
We obtain a dataset including 4,688 task phrases and 6,075 non-task phrases for training a CNN-based task phrase classifier~\cite{Kim2014ConvolutionalNN}. 

Given a task phrase, we extract \textit{Action} and \textit{Object} based on POS tag patterns.
If POS tags match the pattern \textit{Verb+Noun(*)}, we extract \textit{Verb} as Action and \textit{ Noun(*)} as Object, e.g., ``Reads a single character'' is chunked into Action \textit{reads} and Object \textit{a single character}. 
If POS tags match the pattern \textit{Verb+Noun(*)+ADP+Noun(*)} and ``ADP'' means adposition, we extract \textit{Verb+Noun(*)} as Action and \textit{Noun(*)} as Object, e.g., ``Remove element from Collection''  is split into Action \textit{remove element} and Object \textit{Collection}. 

\textbf{b) Task Attribute Extraction.}\label{sec:taskattributeextraction}
Four task attributes will be extracted from the description of a task: 
(1) important \textit{Notes} for the task; (2) \textit{Code Snippet} illustrating how the task works; (3) \textit{Code Summary}; and (4) \textit{API Packet} which is a triplet $<$API name, package (or interface, class) name, the number of parameters$>$. 
For example, \textit{list.add('element')} will be packeted into $<$\textit{add()}, \textit{java.util.List}, \textit{1}$>$. 
Here, we use the number of parameters instead of parameter types for two reasons: 1) matching APIs by the number of parameters is faster than by parameter types (e.g., int, long, float, $<$T$>$, $<$E$>$); and 2) the parameter types identified from text are often not accurate  due to lack of type context.
If multiple notes (or code summaries, code snippets) are extracted for a task, we merge them into one attribute entry.
But we keep each extracted API packet separately as an individual attribute to perform individual API entity linking.

Note that not all task entities have all the four task attributes.
We extract important notes and code summary based on a set of commonly seen keywords for such information in tutorial descriptions, for example, ``if, note, must, remember, instead, only, except, notice'' for important notes, and ``first, second, third, then, finally, this example, in this case for code summary.
Code snippet in programming tutorials are usually enclosed with special HTML tag (e.g, “\textit{$<$codebox$>$}”), which can be easily identified and extracted.

For API packet, we extract the API name mentioned in the sentence with the special HTML tag (e.g, $<$code$>$), for example, \textit{remove()} in ``$<$code$>$remove()$<$/code$>$ element from collection''. 
If there is no special tag for the API name, we use the orthographic feature-based regular expressions \cite{Ye2016LearningTE} (e.g., camel-case style, end with ``()'') to extract API names.
Then we match the API name in the code snippet (if any) to get the corresponding package (or interface, class) name and the number of parameters, which forms the API packet. 
For the matching, we transform the source code into a typed Abstract Syntax Tree involving APIs by using Spoon~\cite{Pawlak2016SPOONAL}, a tool for partial code parsing and type resolution. 
For example, if put() in the task text matches sortedMap.put("a", "one") in the code, we would create an API packet $<$put(), java.util.sortedMap, 2$>$ for the task. 
If the task does not have code snippets or the API name does not match any token in the code snippet, the API packet will have only the API name and the other two slots are null.

\textbf{c) Task Relation Extraction.}
Three types of task relations are extracted: 
(1) \textit{Hierarchical Relation} between a parent task and child tasks; (2) \textit{Sibling Relation} between the same-level child tasks; and (3) \textit{Temporal Relation} for task execution order.
Hierarchical and sibling relations are extracted based on document structure, for example, section and subsection (e.g, HTML tags \textit{$<$h1$>$},\textit{$<$h2$>$}), list bullets (e.g., HTML tags \textit{$<$ul$>$} and\textit{$<$li$>$}). 
Temporal relations are extracted based on commonly used temporal words (e.g, \textit{before}, \textit{after}, \textit{once}) from the task description between the adjacent child tasks.

\vspace{-4mm}
\subsection{KG Fusion}\label{kg-link}
KG fusion is to link the task entities in Task-KG with the API entities in API-KG. 
This is based on API entity linking rather than surface-level keyword matching. 
For example, based on the keyword matching, \textit{add()} mentioned in a task would be confused to be linked to \textit{List.add(E)}, \textit{List.add(int,E)} or \textit{Collection.add(E)}.

\subsubsection{API Representation for Entity Linking}
To achieve high-quality entity linking, we represent an API mention in the task description as a $T_{<AP, Sent>}$ tuple where \textit{AP} is the API packet attribute derived for this API mention and \textit{Sent} is the sentence where this API is mentioned.
We also represent each API entity in API-KG as a $API_{<AP, Sent>}$ tuple where \textit{AP} and \textit{Sent} are the API packet and the function sentence attribute of this API entity.
API packet represents API syntactic structure while the sentence embodies API usage semantics.
Both types of information are needed as several APIs may have the same syntactic structure but their usage semantics would generally be different.
For example, \textit{List.remove(int)} and \textit{List.remove(Object)} have the same syntactic structure (i.e., $<$\textit{remove()}, \textit{List}, \textit{1}$>$), but the former is to remove an element at the given index, while the latter is to remove the given object.

\subsubsection{API Entity Linking}

Entity linking aims to link an entity mention in text to the entity in a knowledge graph.
The entity of our concern is API, and they can be mentioned in task descriptions or user queries.
Without losing the generality, we refer to $Q_{<AP, Sent>}$ as a query tuple (e.g., the tuple for an API mention of a task entity) and \{$API_{<AP, Sent>}$\} as the set of candidate tuples for the API entities in the API-KG.
Note that the API entity linking method introduced here for linking task entities and API entities will also be used for API semantic relation inference (see Section~\ref{sec:enrich API-kg}), task-align relation inference (see Section~\ref{sec:enrich task-kg}), and API/Task-centric knowledge graph search (see Section~\ref{sec:kgapplication}).

First, we use $Q_{<AP>}$ as a query API packet to find a set of candidate API entities in API-KG by matching API syntactic structure (i.e., $API_{AP}$ and $Q_{AP}$).
To check whether two API packets are the same or not, we proposed an API packet matching method.
Given a query $AP_1$$<$x,y,z$>$ and a candidate $AP_2$$<$x',y',z'$>$ where AP$<$x,y,z$>$ represents an API packet$<$API name, package (or interface, class) name, parameter number$>$, if the expression $(x!=null\ \&\&\ x==x') \&\& (y==null \parallel y==y') \&\& (z==null \parallel z==z')$ is true, $AP_1$$<$x,y,z$>$ and $AP_2$$<$x',y',z'$>$ are matched, otherwise the two API packets do not match.
That is, two matched API packets must always have the same non-null API name.
For the package (or interface, class) name (or the parameter number), the matching is relaxed as these two slots of $Q_{AP}$ could be null.
In such cases, the second (or third) slot is regarded as matched.
If the second (or third) slot is not null, then the corresponding slots of the two API packets must be the same to match the two API packets. 

%\subsubsection{Sentence Matching}\label{sentenceMatching}
If API packet matching finds only one API entity, that API entity is linked to the task.
If there are multiple candidate APIs, we filter the candidate API entities by comparing API usage semantics (i.e., $API_{Sent}$ and $Q_{Sent}$).
To check if two sentences are semantically similar, we use the sentence-embedding based matching method~\cite{Ren2020APIMisuseDD,Huang2018APIMR}.

For each sentence, we embed tokens into word embeddings and average all token vectors into a sentence vector. 
We calculate the cosine similarity between the two sentence vectors and select the API entity whose $API_{Sent}$ has the highest cosine similarity with $Q_{Sent}$.
In our current tool, we train the word2vec model~\cite{Mikolov2013DistributedRO} with the 128-dimensional word vectors on all the sentences in the JDK 1.8 API specification and the Java tutorial text.

\vspace{-3.5mm}
\subsection{Enrichment of API Semantic Relations}\label{sec:enrich API-kg}
So far the API-KG contains only API declaration (i.e., syntactic) relations extracted from document structure. 
However, according to our empirical study, APIs often have semantic relations with other APIs.
For example, \textit{Deque.add(E)} and \textit{Deque.offer(E)} supports similar function, but they are also different in that ``if cannot insert the element into Queue, offer() returns false while add() throws exception.''.
Although some API semantic relations can be inferred from API reference documentation~\cite{Liu2020GeneratingCB}, and more types and more fine-grained API semantic relations can be inferred from programming tutorials (i.e., the source of Task-KG).
This is because API reference documentation focuses mainly on describing individual APIs while programming tutorials often explain related APIs from the task perspective.
For example, from Java API specification, existing method~\cite{Liu2020GeneratingCB} infers only a \textit{function-similar} relation between \textit{Deque.add(E)} and \textit{Deque.offer(E)} based on the similarity of their function sentences.
However, from the tutorial about the task ``add an element to the end (tail) of a Deque\footnote{{\href{http://tutorials.jenkov.com/java-collections/deque.html\#add}{http://tutorials.jenkov.com/java-collections/deque.html}}}'', we can infer the subtle behavior-difference (defined in the empirical study~\ref{empir:RQ3}) relation between the two APIs.
According to the definitions of nine categories of API semantic relations in our empirical study~\ref{empir:RQ3},
we develop a pattern-based method to infer these API semantic relations from the task attributes, which greatly enriches API relational knowledge in API-KG.

\begin{table*}
    \centering
    \caption{Sentence Patterns for Extracting API Relation Sentences from Task Descriptions}
    \label{tab:pattern}
    \begin{threeparttable}
    \begin{tabular}{|l|p{5cm}|p{8cm}|}
    \hline
        \cellcolor{mygray}Relation Category & \cellcolor{mygray}Sentence Patterns & \cellcolor{mygray}Example \\\hline
    
    Function Similarity & $AE_1$ [like/similar/same]\ $AE_2$ & 
    \textit{Deque.pop()} is similar to how the \textit{removeFirst()} works\\\hline
    
    Function Opposite & $AE_1$\ opposite (ADP)\ $AE_2$ & The \textit{floor()} does the opposite of the \textit{ceiling()}, \dots\\\hline
    
    Behavior Difference
    & $AE_1$\ and\ $AE_2$\ differ\ in & 
     \makecell[l]{The \textit{add()} and \textit{offer()} methods differ in how the behave\\if the Queue is full so no more elements can be added} \\\hline

    Function Replace & VB\,((ADP)\,NP)\,$AE_1$\,[instead of/rather than/not] $AE_2$ & \makecell[l]{if some of the operations in the transaction fail, \\you would call the \textit{rollback()} instead of \textit{commit()}.}
    \\\hline 

    Function\;Collaboration& $AE_1$\,(be)\,VB(VBN)\,[with/to] $AE_2$ & \textit{DataInputStream} is used with \textit{DataOutputStream} 
    \\\hline
    
    Type Conversion & Convert AE1 to AE2 & Convert \textit{List} to \textit{Set} \\\hline
    
    Implement Constraint & $AE_1$ (will) use $AE_2$ & \makecell[l]{\textit{remove()} method will use
\textit{equals()} to decide\dots }\\\hline

    Logic Constraint & VB\,$AE_1$\,[after/then/until] $AE_2$ & \textit{run()} is executed by the thread after call \textit{start()} \\\hline
        
     Efficiency\,Comparison & $AE_1$\,[fast/easier] than $AE_2$
    &\textit{HashMap} is typically faster than \textit{TreeMap}\dots
    \\
    \hline
    \end{tabular}
    \begin{tablenotes}
    \footnotesize
    \item Note: AE (API entity), VB (verb), ADP (adposition), NP (Noun phrase), ADV (adverb), VBN (past participle).
    \end{tablenotes}
    \end{threeparttable}
    
    \vspace{-0.5cm}
\end{table*}

\vspace{-3.0mm}
\subsubsection{Inference of API Semantic Relations} \label{sec:API-semantic}
To extract API sentences that might contain certain API semantic relations, we summarize 18 sentence patterns from the task description.
Specially, we parse the text description and split the text into sentences with the same software text processing method described in Section~\ref{sec:api-kg}. 
If a sentence starts with a conjunction (e.g, \textit{but}, \textit{and}, etc.), it will not be separated from the previous sentence.
This keeps the relevant sentences as a whole sentence to clarify the context of API semantic relations.
For example, in ``{In order to update the database you need to use a Statement. \textit{But}, instead of calling the executeQuery() method, you call the executeUpdate() method.}\footnote{\href{http://tutorials.jenkov.com/jdbc/update.html}{http://tutorials.jenkov.com/jdbc/update.html}}'', ``in order to update ...'' before ``But'' illustrates when you should use \textit{Statement.executeUpdate(String)} instead of \textit{Statement.executeQuery(String)}.
Then we detect API mentions in the sentences with the API mention extraction method \cite{Ye2016LearningTE}. 
If the number of detected API mentions is greater than two, the sentence will be identified as an API relation sentence. 
Finally, we perform POS tagging for the API sentences with Spacy. 

Three authors collaboratively analyze 10,129 API relation sentences and summarize 18 syntactic patterns for extracting nine types of API semantic relations. Due to space limitations, we only show one of the sentence patterns for each API semantic relation in table~\ref{tab:pattern}. All of 18 sentence patterns can be found in our data package.
If an API sentence matches a pattern, the corresponding API relation will be extracted from the sentence. 
The matching is case-insensitive and the word stem is used for matching different word variants (e.g., \textit{differ*} for \textit{differs}, \textit{different}, \textit{difference}, \textit{differences}). 
Our approach may infer more than one relation for the two APIs from one API sentence, because one API sentence may match several patterns at the same time.
For example, based on the pattern ''$AE_{1}$ [like/same/similar], $AE_{2}$ [except/but]`` or the pattern ``$AE_{1}$ [fast/better/easier/easy] than $AE_{2}$'', we can extract the sentence ``The ConcurrentHashMap is very similar to the java.util.HashTable class, except that ConcurrentHashMap offers better concurrency than HashTable does.''.
At this time, we can extract a \textit{behavior-difference} and an an \textit{efficiency-comparison} relation for \textit{java.util.ConcurrentHashMap} and \textit{java.util.HashTable}.

The API mentioned in an extracted API relation sentence will be linked to the corresponding API entities in API-KG by the API entity linking method described in Section~\ref{kg-link}, and the semantic relation between the two API entities is added to API-KG.
This API relation sentence is also attached to the API entities as an inferred API relation attribute.

\vspace{-4mm}
\subsection{Enrichment of Task Semantic Relations}\label{sec:enrich task-kg}
The Task-KG now contains three types of task relations (parent-child, sibling and temporal) extracted from the tutorial document structure.
However, according to our empirical study~\ref{empir:RQ3}, implicit task semantic relations, such as task-align relation and the task-overlap relation, are often not present in tutorial document structure,
For the former, for example, the tasks ''add elements to List`` and
``add a collection of objects to a Collection'' can accomplish the same goal. 
For the latter, for example, the code for the task ``iterate a list with an iterator'' overlap with the one for another task ``remove elements during iteration''.
In the Java tutorial, these tasks have no explicit relations because they are organized in different sections.
However, with the help of API entities and API relations in API-KG, we can infer task-align and task-overlap relations to enrich Task-KG. 

\vspace{-3.0mm}
\subsubsection{Inference of Task Align Relations}\label{Task-align}
Recall that a task entity is represented as a $<$action, object$>$ tuple.
We infer a task-align relation between the two tasks if the task action phrases are similar and the APIs involved in the object have some relation in API-KG.
For example, as ``add elements'' is similar to ``add a collection of an object'' and ``java.util.list'' extends ``java.util.Collection'', we will infer a task-align relation between ``\textit{add elements to List}'' and ``\textit{add a collection of an object to Collection}''.

Given two task entities ($T_1$ and $T_2$), we calculate a task similarity score for aligning the tasks by $Sim_{task}(T1,T2)=Act_{score}(T1,T2)+Obj_{score}(T1,T2)$ (equation 1). 
If the similarity score is greater than a user-defined threshold, we infer a task-align relation between T$_1$ and T$_2$.
In our current tool, we empirically set the threshold at 1.5 through the experiments on a small validation set of tasks. 
In this equation, $Act_{score}(T1,T2)$ is the cosine similarity between the phrase vectors of the action phrases of $T_1$ and $T_2$.
For the $Obj_{score}(T1,T2)$, we match the object phrase with the API name in the API packets of the task entity to determine if the object mentions an API.
If the Objects of both task entities contain APIs, we transform API mentions into two tuples $<$\textit{AP}, \textit{Sent}$>$ where \textit{AP} is the matched API packet and \textit{Sent} is the task phrase.
We link these two tuples to the API entities in API-KG by the API entity linking method described in Section~\ref{kg-link}.
If the two API entities are linked and there is a certain API relation between the two API entities, $Obj_{score}(T1,T2)$ returns 1; otherwise 0. 
If either one of the Objects does not contain an API, we calculate $Obj_{score}(T1,T2)$ in the same way as $Act_{score}(T1,T2)$ .
 
\subsubsection{Inference of Task Overlap Relations}
We infer task-overlap relations by comparing the code snippets of the two tasks and determining if the code of one task overlaps the partial code of the other in terms of API usage.

Given the code snippet of a task entity, we extract the APIs and transform them into API packets with the API packet extraction method in Section \ref{sec:taskattributeextraction}. 
We retain programming language keywords (e.g, “\textit{while}”, “\textit{if}”) in the code.
Then, we represent the extracted APIs into query tuples $<$\textit{AP},\textit{Sent}$>$ where \textit{AP} is the API packet attribute of this task for an extracted API and \textit{Sent} is the combination of the code summary and note attributes of the task. 
Next we find the corresponding API entities in API-KG through the API entity linking method in Section~\ref{kg-link}.
Finally, we obtain a set of unique API entities used in the task code snippet.
That is, we do not consider the times an API is used.

Given two code snippets ($C_1$ and $C_2$) of the two tasks, we denote two sets of API entities involved in C$_1$ and C$_2$ as $API_{C1}$ and $API_{C2}$, respectively. 
We calculate the code similarity between the two tasks by
$Sim_{code}(C1,C2)=API_{C1} \bigcap API_{C2}/|API_{C1}|$ (equation 2), which computes to what extent $C_1$ overlaps $C_2$ in terms of API usage. 
Note that $Sim_{code}(C_1, C_2)$ could be different from $Sim_{code}(C_2, C_1)$.
So we average $Sim_{code}(C_1, C_2)$ and $Sim_{code}(C_2, C_1)$ as the code similarity of the two tasks. 
If the average score is greater than a given threshold, we deduce a \textit{task-overlap} relation between the two tasks.
In our current tool, we empirically set the threshold at 0.6 through the experiments on a small validation set of tasks.

\vspace{-4mm}
\subsection{Knowledge Graph Application}\label{sec:kgapplication}
Based on our fused and semantically-enriched API-Task KG, we developed a knowledge search tool for recommending programming know-what and know-how knowledge coherently. 
This tool consists of the back-end knowledge graph and the front-end web interface. 
The back-end is a Neo4j
% \footnote{\href{https://neo4j.com/}{https://neo4j.com/}}
graph database which stores our API-Task KG. 
The front-end supports text/code search.
%, but this search is not a simple keyword search.
Instead of simple keyword based search, our search engine uses KG methods to extract task, API and relevant attributes from the query, and perform API/Task-centric search over our API-Task KG.

For example, given the text query ``how to insert an item in List with add()'', the knowledge search results by our tool are shown in Figure~\ref{fig:overview}.
Our search engine extracts a task entity $<$insert an item, List$>$ from the query and uses the task-align method described in Section~\ref{Task-align} to match the task entities in API-Task KG.
Our task-align method can match the relevant tasks even the lexical gap by keywords, for example, ``insert an element at specific index into a list'', and ``insert an item in List''.
In addition to the Best Matched Task, our search engine lists other related tasks through the syntactic and semantic task relations, for example, ``Add element to Set'' which has a task-align relation with the matched task.
Furthermore, our search engine also extracts an API mention \textit{add()} from the query, and links this API mention to the API entity \textit{List.add(int,E}) in API-KG.
As such, it complements the recommended task know-how knowledge with the relevant API know-what knowledge.
Even more, our tool visualizes the KG fragments involving the recommended tasks and APIs from which the user may learn more extended knowledge. For example, through another Task-align relation, the user will learn how to ``Add Collection of Objects to Collection''.

\begin{figure}
    \centering\label{fig:API-centrid}
    \includegraphics[width=0.47\textwidth]{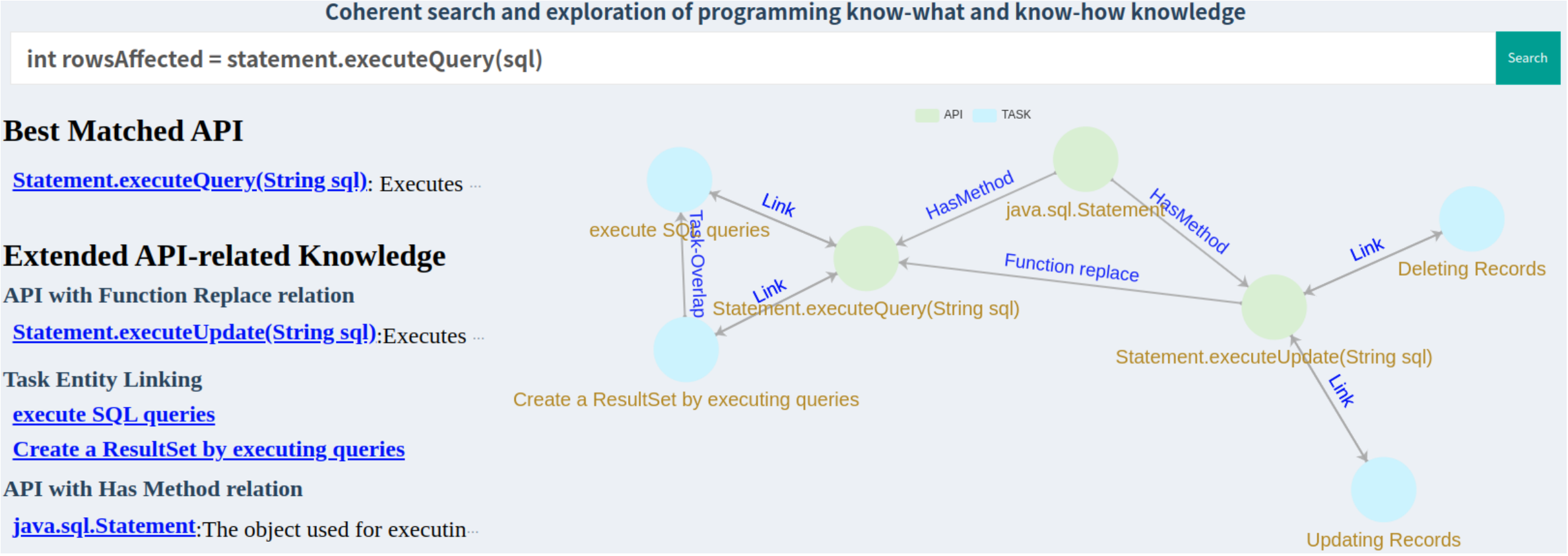}
    \caption{Screenshot of API-centric knowledge search}
    \label{fig:API-centric}
    % \vspace{-0.7cm}
\end{figure}

In a similar vein, given a code query (one or multiple code lines) for an ``execute SQL queries'' task, our search engine extracts the used APIs and constructs the API packets as described in Section~\ref{sec:taskattributeextraction}.
Then, it uses the API packet matching method to find candidate API entities in the API-KG, and lists these API entities as the main API knowledge (see Figure~\ref{fig:API-centric}).
Also, through API-task links, our search engine finds a list of tasks that use the APIs in the code query, and uses the task-overlap method to rank these tasks by their API usage similarities to the code query.
For example, it will recommend related tasks such as ``create a ResultSet by execute queries'' for the code query.
From the knowledge graph fragment visualized on the right panel, the user can learn other extended knowledge, e.g., function-replace relation, the user will learn to substitute \textit{executeUpdate()} for \textit{executeQuery()} when updating the database, and the task ``execute SQL queries'' which can achieve the same goal as the recommended task ``create a ResultSet by executing queries'' (the two tasks have task-overlap relations).

\vspace{-4mm}
\section{Evaluation}
As a proof of concept, we applied our approach to JDK 1.8 API specification\footnote{\href{https://docs.oracle.com/javase/8/docs/api/}{https://docs.oracle.com/javase/8/docs/api/}} and the Java Tutorial\footnote{\href{http://tutorials.jenkov.com/java/index.html}{http://tutorials.jenkov.com/java/index.html}} on the jenkov website.
The resulting API-Task KG
% ~\footnote{\href{https://zenodo.org/record/5709814\#.YZYNtU7iubg}{https://zenodo.org/record/5709721}} 
consists of 8,672 API entities and 7,806 task entities, among which 2,382 API entities and 4,716 task entities are cross-linked.
Our API-Task KG contains 916 API semantic relations and 7,496 task semantic relations which are not available in the existing API-KG~\cite{Li2018ImprovingAC, Liu2019GeneratingQC} and Task-KG~\cite{Sun2019KnowHowIP}.
We conduct a series of experiments to investigate two general research questions: 
\textbf{RQ1} - \textit{What is the accuracy of the key steps of our KG methods and what is the quality of the constructed API-Task KG?}
\textbf{RQ2} - \textit{Can our API-Task-KG-empowered coherent knowledge search answer the frequently-encountered programming questions in practice?}

\vspace{-4mm}
\subsection{KG Quality Evaluation (RQ1)} \label{RQ1}

This RQ examines all our KG steps from API-KG and Task-KG construction, to KG fusion, and enrichment of API and Task semantic relations.
Consider the large amount of data instances in KGs, we adopt a statistical sampling method \cite{Singh1996ElementsOS} that examines the minimal number of data instances so that we can estimate the accuracy of the samples with the error margin 0.05 at 95\% confidence level. 
The sampling size is 384.
We recruit two graduate students (not involved in this work) with two years of Java development experience to label the sampled data instances independently.
We use Cohen's kappa \cite{Landis1977AnAO} to evaluate the inter-rater agreement.
For each sample, if the two students make different labels, a third student is assigned to make an additional label to resolve the conflict by the majority-win strategy.
Based on the final labels, we calculate the data accuracy.

\vspace{-3.0mm}
\subsubsection{API-KG construction}\label{sec:quality-APIKG}

The construction of API skeleton graph is mechanic, relying on only correctly parsing document format and structure.
In our KG, we obtain 44,174 function sentences and 52,685 directive sentences.
The accuracy of these extracted functions and directive sentences need to be examined as they rely on a set of keywords observed by ourselves and previous work~\cite{Martin2011WhatSD}.
We sample 384 function sentences and 384 directive sentences, and ask the annotators to determine whether they are function (or directive) sentences according to the definition in \cite{Liu2019GeneratingQC}.
The Cohen’s kappa between the two students is 0.941 and 0.876 for function sentences and directive sentences, respectively, which indicated almost perfect agreement.
%between the decisions made by the two students. 
Based on the final labels, our keyword-based methods achieve 94.30\% and 92.70\% for the identification of function and directive sentences, respectively. 
The errors are caused by two main reasons.
First, POS tagging errors, e.g., ``peek'' from the sentence ``peek in interface Queue$<$E$>$'' is tagged as a Verb, while it is actually a method name. 
Second, extract incomplete sentences due to incorrect sentences splitting, e.g., ``returns the lowest index i such that (o==null ?''. These errors could be mitigated by software-specific POS tagging and sentence splitting methods. ~\cite{Yitagesu2021AutomaticPT}

\vspace{-3.0mm}
\subsubsection{Task-KG Construction}\label{sec:quality task-kg}

For the task phrase identification, we trained a sentence classifier (TextCNN)~\cite{Kim2014ConvolutionalNN}. 
We compare two training methods: using 10,763 labeled whole sentences (as baseline) or 10,763 labeled verb phrases in these sentences (our method).
The baseline was used in~\cite{Sun2019KnowHowIP}.
Both methods use the same configuration and were evaluated with 10-fold cross-validation. 
The Precision, Recall, F1-Score and Accuracy of our method are 89.79\%, 89.90\%, 89.84\% and 86.56\%, while those of the baseline are 51.36\%, 77.80\%, 61.88\% and 61.36\%. 
Our method performs much better because task phrases are much less noisy than the whole sentences.

For the API-packet attribute extraction, we sampled 384 API packets from 325 task entities, and ask the annotators to determine if all the three slots of the API packets are correct.
Only if all slots are correct, the API packet is regarded as correct.
The two annotators have 0.931 Cohen's kappa (almost perfect agreement).
Based on the final labels, the accuracy of the API-packet attributes is 89.72\%. 
\textbf{The errors come mainly from the inaccurate package (or interface, class) name.} 
For example, the class name of \textit{stream()} should be \textit{java.util.Set}, but it is mistaken as \textit{java.util.Map}.  
This is because, given \textit{Map.keySet().stream()}, we don't know the return type of \textit{Map.keySet()} and thus mistake \textit{java.util.Map} for the class name of \textit{stream()}.
If this is case ``Set s = Map.keySet(); s.stream()'', we can identify \textit{java.util.Set} from the variables.

\vspace{-2.0mm}
\subsubsection{KG Fusion}

For the KG fusion, we sampled 384 API-Task links, and ask the annotators to determine whether the task entity is linked to the correct API entity in API-KG, based on the task entity attributes. 
The two annotators have 0.897 Cohen's kappa (almost perfect agreement). 
Based on the final labels, the entity linking method achieves 86.61\% accuracy for KG fusion. 
Through the error analysis, the errors are mainly caused by inaccurate upper stream API packets, function sentences, code summaries, etc. 
For example, the task ``set a generic type for a Collection'' is incorrectly linked to the ``java.lang.String'', because we inaccurately extract ``Java String‘’ as the API packet $<$\textit{String}, \textit{lang}, \textit{0}$>$ in the task attribute. 
In addition, we also sample 2,400 API entities from API-Task KG and transform these API entities into API-packets. 
86.46\% of API packets are unique. 
This indicates that using the API syntactic structure can distinguish most API entities and thus provide a solid foundation for the high quality fusion of API KG and Task KG.

\subsubsection{Inference of API Semantic Relations}

We examine 916 API semantic relations inferred by our approach.
Table~\ref{tab:apisemanticaccuracy} shows the accuracy of each type of API semantic relation and the Cohen’s Kappa agreement between the two annotators. 
All agreement rates are above 0.76 (at least substantial agreement). 
All accuracies are above 0.95.
The accuracy is 1 for \textit{Function Opposite}, \textit{Type Conversion} and \textit{Efficiency Comparison}, as these three relations generally use clear and specific keywords (e.g., opposite, convert, fast) to match with the syntactic patterns. 
The remaining six API semantic relations are also highly accurate, but two shortcomings: 
1) our syntactic patterns cannot distinguish the APIs from the self-defined methods. 
For example, \textit{stop()} and \textit{dostop()} are extracted as a function-replace relation, but \textit{dostop()} is not an API but a self-defined method.
2) our approach may infer incorrect API relations from API sentences without considering the overall logic of the sentence. 
For example, we infer an API relation $<$\textit{java.util.list}, Logical-Constraint, \textit{java.util.Stack$<$E$>$}$>$ from the API sentence ``each element is \underline{removed} from the \underline{List} then pushed onto the \underline{Stack}'' based on the syntactic pattern \textit{VB AE1 then AE2}. 
However, this sentence expresses only the processing order but not the logical constraint between the two APIs.

\begin{table}
    \centering
    \caption{Accuracy of API Semantic Relation Inference}
    \label{tab:apisemanticaccuracy}
    \begin{tabular}{|m{3.5cm}|m{0.9cm}|m{1.2cm}|m{1.5cm}|}
        \hline
        \textbf{Relation Category} & \textbf{Num} & \textbf{Accuracy} & \textbf{Agreement} \\
        \hline
        Function Opposite & 8 & 1.000 & 1.000 \\
        \hline
        Function Similarity & 116 & 0.965 & 0.893\\
        \hline
        Behavior Difference & 119 & 0.975 & 0.836\\
        \hline
        Function Replace & 95 & 0.958 & 0.821\\
        \hline
        Function Collaboration & 412 & 0.951 & 0.793\\
        \hline
        Implement Constraint & 91 & 0.953 & 0.768\\
        \hline
        Logical Constraint & 43 & 0.952 & 0.780\\
        \hline
        Type Conversion & 23 & 1.000 & 1.000 \\
        \hline
        Efficiency Comparison & 9 & 1.000 & 1.000\\
        \hline
    \end{tabular}
    \vspace{-0.5cm}
\end{table}

\subsubsection{Inference of Task Semantic Relations}

We sample 384 task-align relations and 384 task-overlap relations in our API-Task KG, and ask the annotators to determine their validity.
The accuracy for task-align and task-overlap relations is 88.63\% and 92.47\%, respectively.
The annotators have 0.796 and 0.843 Cohen’s Kappa respectively (almost perfect agreement). 
Through the error analysis, we identify two main causes of errors.
First, the wrong task-align relations are caused by the incorrect task phrases, for example, ``reverse List using Stack'' is task-aligned with ``copy all elements of a List'', but the two tasks have no relation.
For the correctly identified task phrases, the accuracy of inferring task-align relation is actually 97.86\%. 
Second, the wrong task-overlap relations are caused by the incorrect API names extracted from the code. 
For example, our tool may split API names incorrectly because of improper processing of HTML tag (e.g., $<$br$>$). 
As an example, ``stream.$<$br$>$forEach()'' is spilt into two names ``stream'' and ``forEach()'' rather than as a whole. 
Such errors affect the subsequent API entity linking and API usage matching.

%\vspace{1mm}
\noindent\fbox{\begin{minipage}{8.4cm} \emph{Our KG construction, fusion and enrichment steps are accurate and produce a high-quality API-Task KG.} \end{minipage}}\\

\vspace{-4mm}
\subsection{Effectiveness and Usefulness of KG (RQ2)}
We conduct a pilot study to evaluate if our KG-empowered knowledge search could answer the questions on SO.

\vspace{-3mm}
\subsubsection{Dataset}
To simulate real-world programming issues, we select the top 30\% of questions (sorted by the descending order of view counts) from each question topic defined in the empirical study.
All 30 selected questions are listed in Table~\ref{tab:soquestions}.
They had been viewed in total 25,788k times and received 22,308 votes (as of 20th August 2021), which indicates that developers frequently encounter similar problems. 
For each question, we collect the question title and description and the accepted answer.
The descriptions of the last thirteen questions (Q18-Q30) contain problematic code snippets.

\vspace{-3mm}
\subsubsection{Effectiveness Evaluation}\label{3.2.2}

\begin{table}
    \centering
    \caption{30 SO Questions \& Effectiveness Labels}
    \label{tab:soquestions}
    \begin{tabular}{|m{2.8cm}|m{0.7cm}|m{2.8cm}|m{0.7cm}|}
        \hline
        Questions & Label & Questions & Label \\
        \hline
        1.Remove\;equal item from java list. & \ N & 16.Converting\;string to java.util.Date & \ P \\
        \hline
        2.How to Insert an item in List with add()? & \ P & 17.How\;to\;convert java.util.Date to java.sql.Date? & \ P\\
        \hline
        3.Check if a file exists. & \ P & 18.subString()\; and\;subSequence(). & \ P \\
        \hline
        4.Get the first element of the List or Set? & \ N & 19.Difference between\;start() and run(). & \ P \\
        \hline
        5.Efficiently\;iterate entry in Map. & \ P & 20.Read\;the\;value of\;field\;from\;a\;class.  & \ P \\
        \hline
        6.rewrite\;the\;content of\;a file. & \ P & 21.store\;IP\;address range vs location? & \ P \\
        \hline
        \makecell[l]{7.ConcurrentHash-\\Map and Hashtable \\in Java.}  & \ P & 22.check if a date is within a certain range? & \ p \\
        \hline
        8.Sort\;ArrayList\;of custom\;Objects\;by property. & \ N & 23.convert\;an InputStream into a String? & \ P \\
        \hline
        
        9.How\;to\;split\;a string in Java?  & \ N & 24.What\;could cause InvocationTargetException? & \ N \\
        \hline

        10.How\;to\;properly stop the Thread? & \ P & 25.What\,does\,cipher .update\;do in java? & \ P \\
        \hline
        
        11.Avoid\;exception\ when\,removing objects in a loop. & \ P & 26.How do I convert a String to an int in Java? & \ P \\
        \hline
        12.Difference between\;StringBuilder and StringBuffer. & \ P & 27.Difference between add() and offer(). & \ P \\
        \hline
        13.Checking\;file on FTP server. & \ P & 28.Performance\;of TreeMap,\;HasMap?& \ P \\
        \hline
        14.How can I get all the components of\;a\;panel\;in\;Swing?  & \ N & 29.Cannot\;issue data\,statement\,with executeQuery(). & \ P \\
        \hline
        15.start\;a\;function after\;stop\;typing in\;a JTextField?  & \ N & 30.add\;objects\;to List\;throws\;an exception. & \ P \\
        \hline
    \end{tabular}
    \vspace{-0.5cm}
\end{table}

Our tool supports both text and code queries.
For text search, we input the questions' title from Q1 to Q17. 
For code search, we input the code snippets from Q18 to Q30. 
Given a query, our tool outputs the answer including the best matched API or task entity and the extended knowledge of related APIs and/or tasks.
For each question, we use the same graduate students' settings as KG quality evaluation to determine if the answer by our tool includes the accepted answer.
If our answer includes the accepted answer, we regard our tool can positively answer the question (labeled as P), otherwise labeled as negative (N).
We evaluate the inter-rater agreement by Cohen's kappa.
We invite a third student to provide an additional label to resolve the disagreements by majority-win.

Table~\ref{tab:soquestions} presents the analysis results.
For 23 (76.67\%) of the 30 questions (labeled with P), the annotators believe the answers by our tool include the accepted answers of these questions and thus should be able to solve these 23 questions.
Consider Q11 ``Iterating through a Collection, avoiding ConcurrentModificationException when removing objects in a loop?''. 
Given this question title, the text-based search would retrieve the tasks like ``Remove Element From Collection'', ``Iterate a Collection'' and ``Remove Elements During Iteration'' based on the literal text of the task phrases. 
However, our tool recommends the third task as the best match because our tool also considers \textit{ConcurrentModificationException} which is implicit in the description of the third task ``Calling remove() does not cause a ConcurrentModificationException to be thrown.''.

In addition, the inferred API or task semantic relations play an important role in solving the questions.
Take Q25 "What does cipher.update do in java?" as an example.
The user knows how to use \textit{cipher.dofinal()}, but he doesn't know what \textit{cipher.update()} does, 
which often appears with \textit{cipher.dofinal()}. 
Therefore, given a code snippet that contains \textit{cipher.dofinal()},
he hopes search Engine to recommend \textit{cipher.update()}.
Our tool parses \textit{cipher.doFinal()} from the code. Then, based on the inferred Function-Collaboration relation, the tool recommends \textit{cipher.update()} as an extended API which covers the accepted answer. This makes the user understand that \textit{cipher.update()} is used for slicing the large-scale data into small-scale data, and then \textit{cipher.dofinal()} is used for storing the small-scale data.

For the N-labeled questions, our tool only recommends some related knowledge which doesn't directly cover the accepted answer. 
Take Q4 "Get the first element of the List or Set?" as an example.
Given this question, our tool recommends ``Get Elements From a Java List'' as the best match task. 
However, this match is only for \textit{List} but not \textit{Set}.
Then our tool recommends ``Iterate Set Using Iterator'' and ``Iterate List Using Iterator'' as the extended tasks based on the inferred task-align relation.
However, the two extended tasks explain the cursor traversal rather than the index traversal, which does not solve Q4 directly.

\subsubsection{User Study}
This section is to evaluate how our knowledge recommendation helps novice developers solve the 30 SO questions.

\textbf{Baseline}.
To verify the effectiveness of our fused API-Task recommendation, and the helpfulness of the extended suggestions based on knowledge fusion, we use API-KG~\cite{Liu2019GeneratingQC} and Task-KG~\cite{Sun2019KnowHowIP} as our baseline.
API-KG is to declare what an API is, i.e., know-what knowledge. Task-KG is to show how to use an API, i.e., know-how knowledge. The details of API-KG and Task-KG are seen in Section~\ref{sec:Background}.

\textbf{Methodology}. We recruit 12 undergraduate students (not involved in this work) who have novice-level Java development experience. 
Through a pre-study survey, we ensure that none of these students had encountered the experimental questions before.
We assign them into two groups ($G_{A}$ and $G_{B}$) randomly (6 students in each group). 
The $G_{A}$ participants use our tool to answer the questions while those in $G_{B}$ use the independent API-KG and Task-KG tools. 
$G_{B}$ participants can use either or both API-KG and Task-KG tools, but the two tools are separate.

For the $G_{A}$ participants, we give a 20-minute training session to help them learn how to use our tool.
They are asked to perform text search for Q1-Q17 and code search for Q18-Q30.
The $G_{A}$ participants are not allowed to view the information outside our tool.
For the $G_{B}$ participants, we also give them a 20-minute demonstration on how to use API-KG or Task-KG tool, and ask them to search with the single KG tool until the final result is determined.   
It is recommended that, after the participants input the query statement and obtain the query results, they should prefer the query results from SO, and rewrite the original query with the API information or the exception information of the partial code in these results for the next round of the search.
All participants can issue any queries they like, not limited to just question title or description information.
Each question is given eight minutes. 
For each question, the participants are asked to submit an answer in text with the detailed information (not just the reference). 
We also record the process of question answering with the screen recording tool, so that we observe the search behavior of the participants. 
The two authors collaboratively compare the correctness of the participants' answers and collect the completion time of each question per participant.

\begin{table}
    \centering
    \caption{Performance Comparison}
    \label{tab:userstudy}
    \begin{tabular}{|c|c|c|c|}
    \hline
    Groups & Participants & AveQCT (S) & AveAC(\%) \\
    \hline
    \multirow{7}{*}{GroupA} & P1 & 578.7 & 93.33 \\ \cline{2-4} & P2 & 564.6 & 76.67 \\ \cline{2-4} & P3 & 506.8 & 70.00 \\ \cline{2-4} & P4 & 547.9 & 83.33 \\ \cline{2-4} & P5 & 542.8 & 73.33 \\ \cline{2-4} & P6 & 538.9 & 86.67 \\ \cline{2-4}& Ave±stddev & 546.61±22.41 & 80.56±8.03 \\
    \hline
    \multirow{7}{*}{GroupB} & P7 & 780.4 & 66.67 \\ \cline{2-4} & P8 & 713.4 & 56.67 \\ \cline{2-4} & P9 & 753.8.6 & 63.33 \\ \cline{2-4} & P10 & 738.4 & 70.00 \\ \cline{2-4} & P11 & 791.6 & 70.00 \\ \cline{2-4} & P12 & 746.3 & 66.67 \\ \cline{2-4} & Ave±stddev & 753.98±26.02 & 65.56±4.58 \\
    \hline
    \end{tabular}
    \begin{tablenotes}
    \footnotesize
    \item AveQCT: Average Question Completion Time
    \item AveAC: Average Answer Correctness
    \end{tablenotes}
    \vspace{-0.6cm}
\end{table}

\textbf{Results and Analysis.} 
Table~\ref{tab:userstudy} shows the average question completion time (AveQCT) and the average answer correctness (AveAC) for $G_{A}$ and $G_{B}$.
$G_{A}$ complete the questions faster (546.61 vs. 753.98 seconds) and more accurately (80.56\% vs. 65.56\%) than $G_{B}$. 
We use Welch’s T-test~\cite{Bl1947THEGO} for verifying the statistical significance of the differences. 
The difference in time is statistically significant (p$\ll$0.05), while the one in accuracy is insignificant (p=0.01). 

By reviewing the task completion videos, we observed that $G_{B}$ often switched back and forth between the two separate API-KG tool and the Task-KG tool when finding the answer to these questions.
This prolonged their completion time.
In contrast, our KG already distills and organizes the knowledge as explicit API/task entities and their rich relations. Our tool can search and present the relevant knowledge in a well-structured and coherent way.
This helps the $G_{A}$ find the answers faster and more accurately.

Note that the average accuracy of $G_{A}$ is 80.56\%, higher than the 76.67\% effectiveness ratio in our effective evaluation. 
This is because five $G_{A}$ participants inferred the correct answers to three N-labeled questions (Q4/Q8/Q9) based on the extended knowledge even though it does not cover the accepted answers directly.
For example, by reviewing the question-answering videos, we observed that P1
inferred the answer to Q4 by the extended knowledge, such as the ``Iterate the element in List'', and ``Iterate the element in Set'' and learn to get the first element of List or Set with the cursor traversal.
It reveals that the extended knowledge is potentially useful for knowledge acquisition.

$G_{A}$ had higher accuracy on ten questions (Q6/Q12/Q-\\18/Q19/Q20/Q24/Q25/Q26/Q28/Q29), compared with $G_{B}$.
Specifically, for Q6/Q7/Q12/Q18/Q19/Q28, five $G_{A}$ participants answered correctly while only two $G_{B}$ participants answered correctly.
% five $G_{A}$ participants answered the two ways to rewrite the file while only two in $G_{B}$ answered the two ways. 
For example, through the videos, in order to solve Q6, we noticed that P3 and P9 used the same query for retrieval(i.e., ``How to rewrite the content of a file''). The difference is that P3 directly acquired the correct answer through the \textit{Task-align} relation between the two tasks given by our tool, while P9  got only one way to rewrite files from Task-KG tool that did not satisfy the question answer.
For Q20/Q24/Q25/Q26/Q29, six $G_{A}$ participants answered correctly while none of them in $G_{B}$ were answered correctly.
For example, through the question-answering videos, in order to solve the Q29, we observed that P11 first input the misused method (i.e., \textit{java.sql.Statement.executeQuery()}) as the query to search with the API-KG tool, and got the \textit{java.sql.Statement.executeQuery()} definition from the searching results but he failed to fix the programming issues. 
And then P11 reformulated the query with the exception information thrown by the program (i.e., \textit{can not issue data manipulation statements with executeQuery()}), and inputted it into the Task-KG tool. 
However, he only got the task ``how to use executeQuery() execute the statement in SQL.'',  and fail to fix the programming issues again.
And the P6 inputted the issue code statement into our tools, and acquired the answer through the \textit{Function-Replace} relation between \textit{Statement.executeQuery(String)} and \textit{Statement.executeUpdate(String)} given by our tool and he succeeded in fixing the programming issue.

Both $G_{A}$ and $G_{B}$ had higher accuracy (78.24\% average accuracy for $G_{A}$ vs. 71.62\% for GB) on eighteen questions (Q2/Q3/Q4/Q8-Q11/Q13-Q17/Q21/Q22/Q23/Q27/Q30),
but $G_{A}$ spent on average 42.67\% less time than GB, especially on Q11 and Q12, $G_{A}$ was 51.09\% faster. 
For example, through the task completion videos, we observed that P7
in $G_{B}$ learned the answer to Q27 by reading the definition of
\textit{Deque.add()} and \textit{Deque.offer()} in different search results given by the API-KG tool, while P1 in $G_{A}$ directly acquired the answer through the \textit{function-similar} and \textit{behavior-difference} relations between the two APIs given by our tool. 
It reveals that the semantic knowledge generated by our fused KG can help to solve the questions that cannot be solved by
API-KG or Task-KG independently.

Both $G_A$ and $G_B$ had low accuracy on Q1 and Q5 (41.67\% for $G_A$ and 33.33\% for $G_B$). 
For Q1, we observed that P7 in $G_{B}$ got the task ``how to use remove() in the list'' from the Task-KG tool, but he did not get a cue for overriding \textit{equals()}.
After some investigation, P3 in $G_{A}$ noticed the logical-constraint between \textit{remove()} and \textit{equals()} recommended by our tool, but she ran out of time to submit the answer.
For Q5, P5 in $G_{A}$ learned three ways to iterate entry in Map with our tool, i.e., by using \textit{Iterator}, by using \textit{foreach}, and by using Lambda Expression. At this time,  because our tool currently does not extract efficiency-comparison between the three ways to accomplish the task. P5 randomly copied one of the ways as the answer without carefully examining the differences between the three ways, which was an incorrect answer to Q5.
P7 in $G_{B}$ only got one way from the Task-KG tool, which was incorrect.

\begin{figure}
    \centering
    \includegraphics[width=0.47\textwidth]{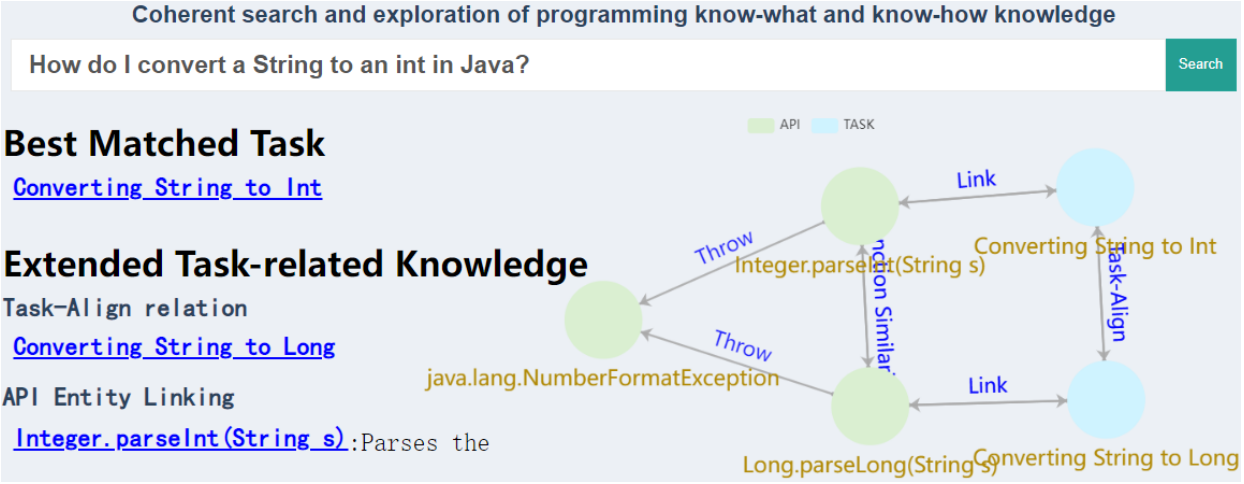}
    \caption{Screenshot of the Q26 search results}
    \label{fig:convert_example}
    \vspace{-0.6cm}
\end{figure}

% First, fusing know-what and know-how knowledge enables text/code-centric search that is impossible with API (or Task) KG.
In summary, compared with the API- or Task-KG, our proposed API-Task KG can improve the efficiency and accuracy of developers in completing programming tasks. 
The API-Task KG has two advantages. 
First, it bridges the gap between know-what and know-how knowledge. 
As a result of the knowledge fusion, our tool provides developers with not only the definition and caveats about the API~\cite{Li2018ImprovingAC}, but also the usage of the API~\cite{Sun2019KnowHowIP}.
Take Q26 for example.
Five $G_{A}$ participants used a ``try-catch'' structure to optimize the code example in the task returned by the search engine.
Figure~\ref{fig:convert_example} shows P4's search result for Q26, in which the Throw relation between \textit{Integer.parseInt()} and \textit{java.lang.NumberFormatException} prompts him to get the correct answer.
In contrast, none of $G_{B}$ participants answered correctly due to the knowledge gap.
Second, the API-Task KG contains eleven semantic relations that are not present in the stand-alone API-KG or Task-KG. 
These semantic relations play a very important role in solving programming tasks.
Q28 illustrates that Efficiency-Comparison helps improve programming with a more efficient API; Q29 illustrates that Function-replace helps fix code bugs.

\noindent\fbox{
\begin{minipage}{8.4cm} \emph{Our evaluation provides the initial evidence of the effectiveness and usefulness of our KG methods for answering programming questions that developers frequently encounter. 
} \end{minipage}}
\vspace{-3pt}

\vspace{-2.5mm}
\section{DISCUSSION} 
The major threat to internal validity is the subjective judgment of the sampled data instances and the answer correctness. 
To mitigate this issue, we invited the annotators who were not involved in our work.
They annotated the data independently and had high inter-rater agreements.
We make our experimental data available for replication, which can be further validated the community.
One major threat to external validity is our approach was evaluated on only Java documentation.
Another external threat is the limitation of the NLP-based tools. For example, when performing POS tag, Spacy may mistake the API name as ``VERB'' or other parts of speech. For another example, when Spacy splits a line text into multiple sentences, the sentence split error may occur due to the unexpected HTML tag(e.g., $<$br$>$).
Moreover, an external threat is our user study involves only 30 programming questions.
These questions are common programming issues, covering four different aspects, task-oriented (e.g., Q3/Q5), API comparison (e.g., Q2/Q12/Q18/Q27/Q28), code optimization (e.g., Q1/Q25) and code debugging (e.g., Q29/Q30).

For the API-Task KG construction, we construct API-KG using the methods mentioned in~\cite{Li2018ImprovingAC, Liu2019GeneratingQC}, and Task-KG using the method mentioned in~\cite{Sun2019KnowHowIP}.
And then we fuse API-KG and Task-KG based on the API entity linking.
However, when extracting API entities from code examples in Task-KG, the package name in the extracted API packet may be lost due to the lack of import statements in the code examples.
We will infer the package name of the API using method~\cite{saifullah2019learning} to further improve the accuracy of entity linking.
For the inference of API semantic relations, we extract API relation sentences from programming documents mentioned in section~\ref{sec:API-semantic} using the heuristic rules.
However, this method lacks the generality when faced with different programming documents of the different styles.
We will enhance the generality of extracting entities (or relations) by using a large language model, e.g., UIE framework~\cite{lu2022unified}.

Our proposed approach is of great significance.
For the semantic relations inferred, 1) they make search results interpretable.
For example, for Q29, our tool recommends \textit{statement.ExecuteQuery(String)} as the best result, and \textit{Statement.executeUpdate(String)} as the extended result.
The Function-replace relation between \textit{Statement.ExecuteQuery(String)} and \textit{Statement.executeUpdate(String)} can explain why such search results are obtained.
2) The semantic relations help recognize more development needs.
API comparison is defined in~\cite{liu2021api} and used in search engines to recognize development needs.
By contrast, our proposed approach provides nine API semantic relations.
For example, for Q7, the best search answer reveals not only the Behavior-difference relation between \textit{java.util.ConcurrentHashMap} and \textit{java.util.HashTable}, but also their Efficiency-comparison relation.
So many API semantic relations between these two API entities can help recognize development needs exactly.
For the resulting API-Task KG, multiple applications can benefit from it.
1) It extends the code search in a better space. The API entities of our KG can be used to match the under-development code to provide more code examples for the developers;
2) Our fused KG provides the interoperability for the code debugging. For the misused API involved in the exception information, it tells how to fix this error.
3) Our tool is also useful for code optimization. Through analyzing the Efficiency Comparison relation, we can replace the current API with the faster API.

\vspace{-4mm}
\section{Related Work}

Software documentation is a common way to communicate programming knowledge~\cite{Martin2011WhatSD}.
Several studies identify the challenges in accessing and using the knowledge in documentation~\cite{Martin2011WhatSD}.
% \cite{https://link.springer.com/article/10.1007/s10664-011-9186-4, ??martinrobillardhasapaperonthis}.
Recently, knowledge graph methods have been proposed to improve the knowledge accessibility by lifting the knowledge from document text to explicit APIs, tasks, and their relations and constraints~\cite{Li2018ImprovingAC, Liu2019GeneratingQC, Liu2020GeneratingCB, Ren2020APIMisuseDD, Sun2019KnowHowIP, Sun2020TaskOrientedAU}.
For example, Li et al. \cite{Li2018ImprovingAC} and Liu et al. \cite{Liu2019GeneratingQC} construct API KG with API caveats or directives.
Such API-KGs support API-centric caveat search or knowledge summary.
Sun et al. \cite{Sun2019KnowHowIP} construct a task-oriented KG and support task-centric search.
They further enhance the Task-KG with more actions and fine-grained code snippets, and integrate the Task-KG into the IDE to support task-aware API recommendation\cite{Sun2020TaskOrientedAU}.

All these works focus on either API (know-what) knowledge or task (know-how) knowledge.
Liu et al. \cite{Liu2020GeneratingCB, Liu2019GeneratingQC} fuse API knowledge and general computing knowledge from Wikipedia, but our work is the first attempt to fuse the two types of programming knowledge.
In addition, our approach enriches the fused knowledge graph with fine-grained API and task semantic relations.
Liu et al.~\cite{Liu2020GeneratingCB} infer three types of API comparison relations (function, constraint and efficiency), which are similar to our API semantic relations but much coarse grained.
Ren et al.~\cite{Ren2020APIMisuseDD} converts API caveat sentences into explicit API constraint relations which supports API misuse detection.
Their constraints correspond to only one type of our API semantic relations (i.e., logical constraint). 
Some studies~\cite{Liu2020GeneratingCB,Zhong2009InferringRS,Tan2007icommentBO,Zhou2017AnalyzingAD} infer logical constraints from text, but they express the constraints as logic formulas rather than KG.
The formulas-supported constraint is a hard match limited to a few specific sentence patterns, while the KG-supported one is a soft match with the strong generalization ability of being suitable for more sentence patterns.
This is because we only need to consider entities and relations, not their order.

In addition to software documentation, much programming knowledge exists in heterogeneous formats such as code, forum discussions, programming screencasts.
Recovering the traceability across these heterogeneous formats receives much attention.
Srinivas et al.~\cite{Srinivas2020Graph4CodeAM} build Graph4Code by connecting source code to various usage documentation to enrich program code semantics, while Bacchelli et al.~\cite{bacchelli2010linking} link the source code to informal text in email.
Treude et al.~\cite{Treude2016AugmentingAD} augment API documentation with insight sentences on SO using a machine learning-based approach.
All of these works link heterogeneous knowledge by only comparing sentence semantics at the text level, which might lead to confusion when linking two entities because some APIs may have the same description.
For instance, \textit{java.util.List.get(int index)}, \textit{java.util.ArrayList.get(int index)}, and \textit{java.util.LinkedList.get(int index)} all have the same description: ``Returns the element at the specified position in this list''.
At this point, the task entity is unsure which API entity should be linked.
To address this issue, studies~\cite{dagenais2012recovering,Subramanian2014LiveAD} link API and its learning resources (e.g., tutorial) based on the API fully-qualified name (i.e., API syntactic structure).
The API fully-qualified name, however, cannot be correctly inferred because the code example is always syntactically incomplete~\cite{Huang2022PrompttunedCL}.
In contrast, we devise an entity-based approach to knowledge fuse that compares the Task-KG task entity to the API-KG API entity while taking both the API syntactic structure and usage semantic into account.
In addition, studies~\cite{Petrosyan2015DiscoveringIE,Treude2016AugmentingAD,Ren2020APIMisuseDD} show that the complementarity of different information sources can present a complete picture of relevant knowledge.
With API and task as explicit entities in a graph, our approach supports coherent API- and task-centric knowledge search and presents programming know-what and know-how knowledge as a unified whole.

\vspace{-3mm}
\section{Conclusion}
In this paper, we conduct an empirical study on the API usage questions on Stack Overflow, and identify the necessity and mutual enrichment of programming know-what and know-how knowledge. Based on this study, we present a novel approach for constructing a fused and semantically enriched knowledge graph of APIs and tasks. Supported by this knowledge graph, our knowledge search engine achieves the effect of ``1+1$>$2'' through coherent API- and task-centric matching and recommendation.

\vspace{-3mm}
\section{Acknowledgements}
The work is partly supported by the National Nature Science Foundation of China under Grant (Nos. 61902162, 61862033, 62262031), the Nature Science Foundation of Jiangxi Province (20202BAB202015), and Postgraduate Innovation Fund Project of Jiangxi Province(YC2021-S308).

\bibliography{sample-base}

\par\noindent 
\parbox[t]{\linewidth}{
\noindent\parpic{\includegraphics[height=3.0in,width=1in,clip,keepaspectratio]{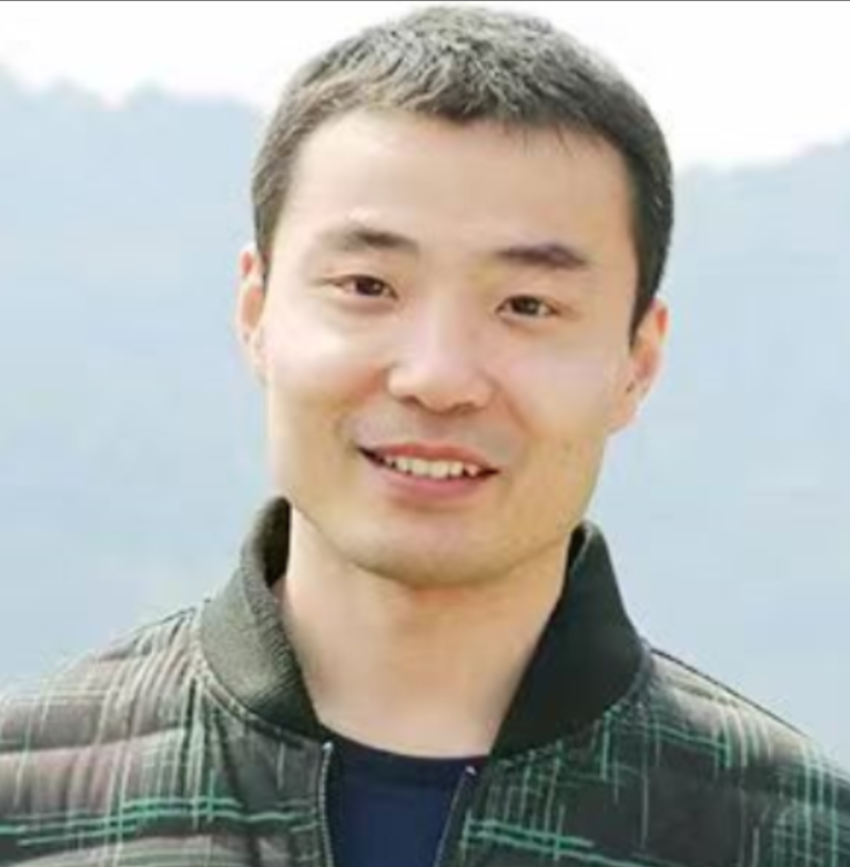}}
\noindent {\bf QING HUANG}\
received the M.S degree in computer application and technology from Nanchang University, in 2009, and the PH.D. degree in computer software and theory from Wuhan University, in 2018. He is currently an Assistant Professor with the School of Computer and Information Engineering, Jiangxi Normal University, China. His research interests include information security, software engineering and knowledge graph.}
% \vspace{0.2\baselineskip}

\par\noindent 
\parbox[t]{\linewidth}{
\noindent\parpic{\includegraphics[height=3.0in,width=1in,clip,keepaspectratio]{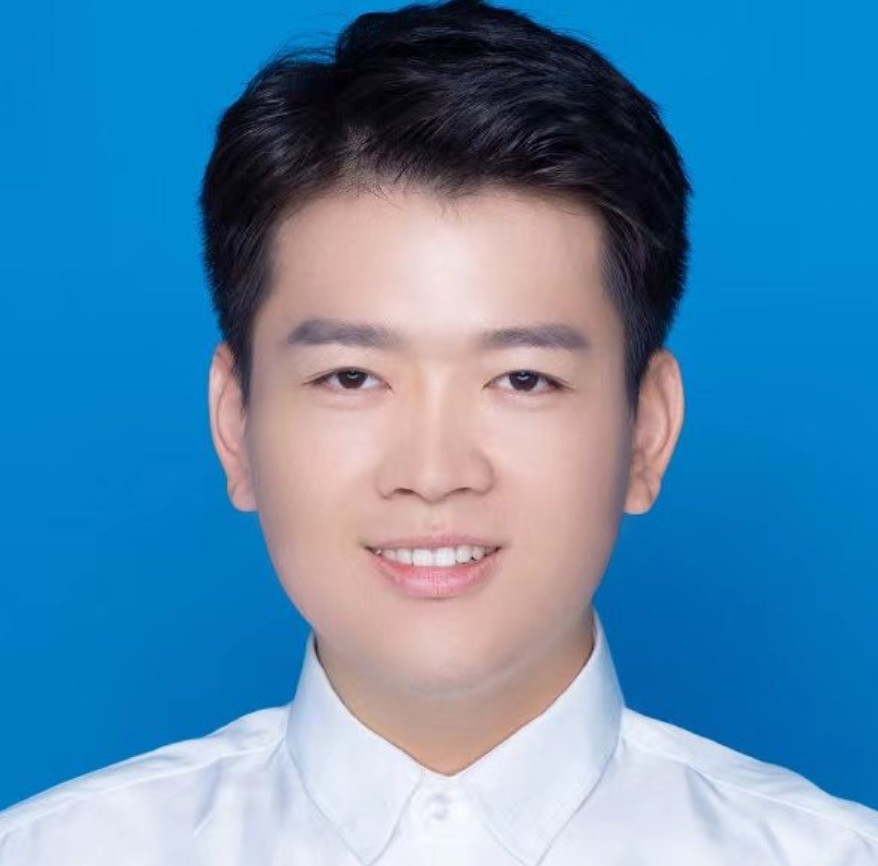}}
\noindent {\bf Zhiqiang Yuan}\
is a second-year graduate student in the School of Computer and Information Engineering, Jiangxi Normal University. His research interests are software engineering and knowledge graph.
}
% \vspace{1\baselineskip}

\par\noindent 
\parbox[t]{\linewidth}{
\noindent\parpic{\includegraphics[height=3.0in,width=1in,clip,keepaspectratio]{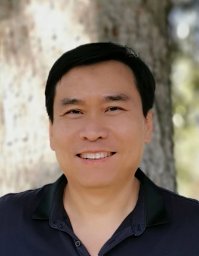}}
\noindent {\bf Zhenchang Xing}\
is an Associate Professor in the Research School of Computer Science, Australian National University. Previously, he was an Assistant Professor in the School of Computer Science and Engineering, Nanyang Technological University, Singapore, from 2012-2016. His main research areas are software engineering, applied data analytics, and human-computer interaction.}

\par\noindent 
\parbox[t]{\linewidth}{
\noindent\parpic{\includegraphics[height=3.0in,width=1in,clip,keepaspectratio]{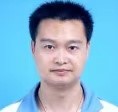}}
\noindent {\bf ZHENGKANG ZUO}\
received the Ph.D. degree in computer science and technology from the Chinese Academy of Sciences (CAS), Beijing, China. He is currently a professor and deputy director of the Computer Science and Technology Department of Jiangxi Normal University, Nanchang, China. His main research interests include software formal methods, generic programming, etc.}
% \vspace{4\baselineskip}

\par\noindent 
\parbox[t]{\linewidth}{
\noindent\parpic{\includegraphics[height=3.0in,width=1in,clip,keepaspectratio]{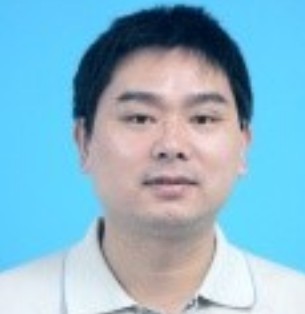}}
\noindent {\bf Changjing Wang}\
received the bachelor's and master's degrees from Jiangxi Normal University, China, in 1999 and 2004, respectively. He received the PhD degree from Institute of Software, Chinese Academy of Science, China, in 2012. He is currently a professor with the College of Computer Information and Engineering, Jiangxi Normal University, China. His research interests include Web service and formal method.}
\vspace{0.5\baselineskip}

\par\noindent 
\parbox[t]{\linewidth}{
\noindent\parpic{\includegraphics[height=3.0in,width=1in,clip,keepaspectratio]{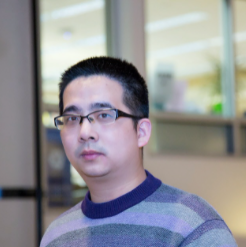}}
\noindent {\bf Xin Xia}\
is the director of the Software Engineering Application Technology Lab at Huawei, China. He received the \textbf{ACM SIGSOFT Early Career Researcher Award} in 2022. His current research focuses on data science for software engineering, i.e., mining and analyzing rich data in software repositories to uncover interesting and actionable information.}

\end{document}